\documentclass[journal,10pt,letterpaper,english]{IEEEtran}

\usepackage{babel}
\usepackage[utf8x]{inputenc}
\usepackage{graphicx}
\usepackage{amsmath}
\usepackage{microtype}
\usepackage{color}
\usepackage{microtype}
\usepackage{cite}
\usepackage{tikz}

\newcommand\copyrighttext{%
	  \footnotesize \textcopyright 2017 IEEE. Personal use of this material is permitted.
	    Permission from IEEE must be obtained for all other uses, in any current or future 
	      media, including reprinting/republishing this material for advertising or promotional 
	        purposes, creating new collective works, for resale or redistribution to servers or 
		  lists, or reuse of any copyrighted component of this work in other works. 
		    }
		    \newcommand\copyrightnotice{%
			    \begin{tikzpicture}[remember picture,overlay]
				    \node[anchor=south,yshift=10pt] at (current page.south) {\fbox{\parbox{\dimexpr\textwidth-\fboxsep-\fboxrule\relax}{\copyrighttext}}};
			    \end{tikzpicture}%
			    }

\begin{document}

\title{Delay Properties of Energy Efficient Ethernet Networks}

 \author{Miguel~Rodríguez-Pérez,~\IEEEmembership{Member,~IEEE,} %
   Sergio Herrería-Alonso,\\%
   Manuel Fernández-Veiga,~\IEEEmembership{Senior Member,~IEEE,} %
   and~Cándido~López-García%
   \thanks{The authors are with the Telematics Engineering Dept., Univ. of
     Vigo, 36310~Vigo, Spain. Tel.:+34~986813459; email: miguel@det.uvigo.gal}}
% \author{Author 1, %
%   Author 2,\\%
%   Author 3 %
%   and~Author 4%
%   \thanks{The authors are with the ---the affiliation has been redacted to
%     comply with the double-blind review procedure.---}}

\maketitle
\copyrightnotice

\begin{abstract}
  % 75 -- 100 words
  Networking operational costs and environmental concerns have lately driven
  the quest for energy efficient equipment. In wired networks, energy
  efficient Ethernet (EEE) interfaces can greatly reduce power demands when
  compared to regular Ethernet interfaces. Their power saving capabilities
  have been studied and modeled in many research articles in the last few
  years, together with their effects on traffic delay. However, to this date,
  all articles have considered them in isolation instead of as part of a
  network of EEE interfaces. In this paper we develop a model for the traffic
  delay on a network of EEE interfaces. We prove that, whatever the network
  topology, the per interface delay increment due to the power savings
  capabilities is bounded and, in most scenarios, negligible. This confirms
  that EEE interfaces can be used in all but the most delay constrained
  scenarios to save considerable amounts of power.
\end{abstract}

% \begin{keywords}
%   Energy efficiency, power awareness, Ethernet, IEEE 802.3az.
% \end{keywords}

\IEEEpeerreviewmaketitle

\section{Introduction}
\label{sec:introduction}

\IEEEPARstart{A}{s part} of the ongoing effort to reduce energy demands of
networking infrastructure, energy efficient Ethernet (EEE) interfaces were
first standardized in~\cite{802.3az} for transmission speeds up to 10~Gb$/$s in
copper medium, and later expanded to 40 and 100~Gb$/$s speeds
in~\cite{IEEE802.3bj}. Their introduction has permitted significant energy
usage reductions of up to 90\% in low load periods with little additional
delay. % These energy efficient interfaces have
% the capacity to enter low power modes to save energy when there is no
% traffic to be send. The standards do not regulate how an interface
% should decide when to enter (or exit) the low power modes,
EEE interfaces of speeds up to 10~Gb$/$s have one low power idle (LPI) mode
designed to save energy when there are no transmissions. Usually, an EEE
interface transitions to this LPI mode when the transmission queue depletes.
To minimize latency,
%, the frame transmission
%algorithm~\cite{herreria12:_gi_g_model_gb_energ_effic_ether} is used:
the normal operating mode is restored as soon as a new frame is ready for
transmission~\cite{reviriego11:_initial_evaluat_energ_effic_ether}. The
transitions to the LPI mode and back take a non negligible amount of time (of
the order of a single frame transmission) that increases latency and wastes
some energy, as the energy needs are comparable to that of an active
interface.

Several articles have studied and modeled the performance of EEE
interfaces~\cite{herreria12:_gi_g_model_gb_energ_effic_ether,larrabeiti11:_towar_gb_ether,reviriego11:_initial_evaluat_energ_effic_ether,Cenedese2016}.
However, to the best of our knowledge, all of them have considered the
simplified case of an isolated switch. This article models the effects of a
network of 10~Gb/s EEE interfaces on the traffic delay. We consider two
representative network topologies that form the basis of more complex ones. In
the first one, a single EEE interface aggregates the traffic coming from a
group of Ethernet interfaces. The second one consists on a series of EEE
interfaces. We have found that the aggregating interface behaves like an
isolated EEE one, irrespective of whether the previous interfaces in the
network have energy saving capabilities. In the second case, we demonstrate
that the per interface added delay converges to a constant value. Finally, we
prove that, in both cases, the penalty for deploying EEE interfaces, i.e. the
delay increment per interface compared to regular Ethernet interfaces, is
bounded.

\section{Traffic Aggregation}
\label{sec:concentrators}

\begin{figure}
  \centering
  \includegraphics[width=.912\columnwidth]{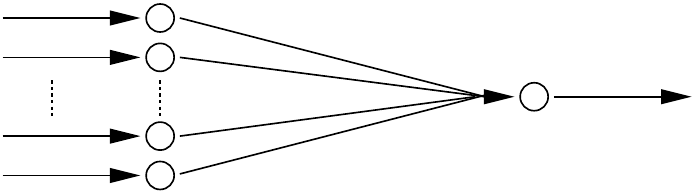}
  \caption{A group of interfaces sending traffic to an
    aggregating switch.}
  \label{fig:aggregating-net}
\end{figure}
We first consider a two stage network like the one shown in
Fig.~\ref{fig:aggregating-net}. At the first stage, a group of interfaces send
traffic to a single switch at the second stage. This second switch aggregates
traffic to a single outgoing link via an EEE interface. We are interested
in the delay suffered at this second stage.

There already exist several results in the literature for the delay model of
an isolated energy efficient
interface~\cite{herreria12:_gi_g_model_gb_energ_effic_ether,Mostowfi2012,Kim2013,Bolla201416}.
However, most of them require the frame arrivals to form a Poisson (or
Poisson-like) process. Unfortunately, the output of the interfaces in the
first stage hardly ever follows a Poisson process. This only happens if the
frame sizes follow an exponential distribution, the arrivals already followed
a Poisson process at the first stage and the first stage interfaces have no
vacations, i.e., they are not energy efficient themselves~\cite{Burke1956}.

However, when we consider the aggregated output of the first stage as a whole,
we can make use of the Palm–Khintchine theorem that states that this aggregate
resembles a Poisson process when the number of independent contributors is
large enough. We make use of this result, and of the average delay model for
Poisson traffic in~\cite{herreria12:_gi_g_model_gb_energ_effic_ether}
($\overline W_{\text{EEE}}$), to approximate the average delay at the second stage
interface as
\begin{equation}
  \label{eq:n-whatever}
  \overline W_{\text{agg}} \approx \overline W_{\text{EEE}} = \frac{1+\lambda^2\sigma_S^2}{2\lambda(1-\rho)}+
  \frac{1-\rho}{2\lambda}+
  \frac{(\lambda T_{\mathrm{W}})^2-2}{2\lambda(1+\lambda T_{\mathrm{W}})},
\end{equation}
where $\lambda$ is the average incoming traffic rate, $\sigma_S^2$ is the
transmission time variance, $\rho$ is the traffic load and $T_{\mathrm{W}}$ is
the setup time of the interface.

However, the delay in~\eqref{eq:n-whatever} is not only caused by the energy
saving algorithm. A significant part is due to regular traffic queuing.
According to the Pollaczek-Khinchine formula, the queuing delay at a regular
Ethernet interface receiving Poisson traffic can be calculated as
\begin{equation}
\label{eq:Wmg1}
  \overline W_{\text{M/G/1}}= \frac{\rho+\lambda \mu \sigma_S^2}{2(\mu-\lambda)},
\end{equation}
with $\mu=1/\overline S$, and $\overline S$ denoting the average transmission
time.

Therefore, the delay caused by the energy savings capabilities is clearly
\begin{equation}
  \label{eq:delta-wagg-1}
  \begin{split}
  \Delta \overline W_{\text{agg}} &= \overline W_{\text{agg}} - \overline
  W_{\text{M/G/1}} \\
  &= \frac{1-\rho^2}{2\lambda(1-\rho)} +
  \frac{1-\rho}{2\lambda}+
  \frac{(\lambda T_{\mathrm{W}})^2-2}{2\lambda(1+\lambda T_{\mathrm{W}})},  
\end{split}
\end{equation}
that after some straightforward algebra becomes
\begin{equation}
  \label{eq:delta-wagg-final}
  \Delta \overline W_{\text{agg}} = \frac{T_{\mathrm{W}}}2\left(
    1+\frac{1}{1+\lambda T_{\mathrm{W}}}
  \right).
\end{equation}

Note that
$T_{\mathrm{W}}/2 \le \Delta \overline W_{\text{agg}} \le
T_{\mathrm{W}}\,\,\forall \lambda \ge 0$ and that, at the same time, it does not
have any dependence on the variance of the frame sizes and just depends on the
average arrival rate.

\section{Interfaces in Tandem}
\label{sec:tandem-links}

\begin{figure}
  \centering
  \resizebox{.75\columnwidth}{!}{\input{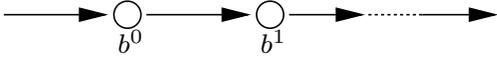}}
  \caption{A network with two interfaces in tandem.}
  \label{fig:tandem-net}
\end{figure}
We now consider a network composed of two network interfaces in tandem ($b^0$
and $b^1$) as depicted in Fig.~\ref{fig:tandem-net}. Again, we are interested
in the waiting time at the second interface, as the delay at the first one has
already been studied in the literature. Although it is infrequent to encounter
in practice a series of network interfaces, it is a valid approximation to the
case where a single incoming port of a switch represents the majority of the
incoming traffic, even when the main contributor changes over time from one
port to another.

For the analysis we will consider two different cases: one where the first
interface is not energy efficient, i.e., it is a regular Ethernet interface,
and a second one where both interfaces are energy efficient. In both cases we
assume that all links have the same nominal capacity.

\subsection{A Regular Ethernet Interface Followed by an EEE Interface}
\label{sec:non-power-aware}

We now consider the case of a regular Ethernet interface followed by an EEE
interface. This system, when considered as a whole, can be directly modeled as
a single EEE interface. The first interface shapes the traffic reaching the
second interface as if it were a token bucket with generating rate equal to
the outgoing link capacity. For constant size frames, it is easily seen that
the interarrival times of the frames reaching the second interface are never
shorter than the transmission time. Hence, any queue at the second interface
must only be due to its energy saving algorithm, and must remain otherwise
unaffected by the actual traffic load, as it is capped at the first
interface.\footnote{For variable frame sizes, a small queue is formed, albeit
  limited to the maximum difference of frame sizes. When a frame shorter than
  the first one of the current busy cycle arrives, it stays in the queue
  during a time equivalent to the difference between the length of their
  transmission times.} The average waiting time at the second interface in
this case can be calculated subtracting the waiting time at a regular Ethernet
interface ($\overline W_{\text{regular}}$) from the waiting time of an energy
efficient interface ($\overline W_{\text{EEE}}$):
\begin{equation}
  \label{eq:droptail-frame-1}
  \overline W_{\text{tandem}}^{\text{regular}} = \overline W_{\text{EEE}} - \overline W_{\text{regular}}.
\end{equation}

In the case where the arrivals at the first interface form a Poisson process,
$\overline W_{\text{EEE}} = \overline W_{\text{agg}}$
from~\eqref{eq:n-whatever} and
$\overline W_{\text{regular}} = \overline W_{\text{M/G/1}}$.
So, it is clear that 
\begin{equation}
  \label{eq:droptail-frame-1-final}
  \overline W_{\text{tandem}}^{\text{regular}} = \Delta \overline
  W_{\text{agg}} = \frac{T_{\mathrm{W}}}2\left(
    1+\frac{1}{1+\lambda T_{\mathrm{W}}}
  \right).
\end{equation}

In this case, all the delay is due to the energy saving algorithm. Hence, as
before, this delay is bounded by $T_{\mathrm{W}}$ and is only dependent on the
average traffic rate.

\subsection{A Series of Interfaces in Tandem}
\label{sec:two-power-aware}

A more interesting scenario is the one formed by two (or more) consecutive
energy efficient interfaces. Although the result will be surprisingly simple,
the model is a bit more elaborated than the previous one.

According to the status of the first interface ($b^0$), a new frame can arrive
to it either while it is \emph{sleeping}, \emph{transitioning to sleep},
\emph{active} or \emph{transitioning to active}. We first consider the case
where the first interface is either \emph{active} or \emph{transitioning to
  active} when a frame arrives. Let frame $i$ be the $i$-th frame to arrive at
the interface in the current busy cycle, so it either arrives at $b^0$ while
it is \emph{active} or \emph{transitioning to active.} Then, according to the
Lindley's recursion, its waiting time at the second interface $b^1$ is
$W^1_i = W^1_{i-1} + s^1_{i-1} - x^1_i$, with $s^j_{i-1}$ being the service
time of frame $i-1$ at interface $b^j$ and $x^j_i$ the interarrival time at
$b^j$ between frames $i-1$ and $i$. However, it is clear that $x^1_i = s^0_i$,
as the queue at $b^0$ was, by hypothesis, non empty when $i$ arrived at it. At
the same time, $s^0_i = s^1_i$ as the capacity of both links attached to the
interfaces is the same, so we must conclude that
$W_i^1 = W_{i-1}^1 + s^1_{i-1}-s^1_i$. Finally, recall that frame 1 must have
reached $b^1$ in the sleeping state, as it is the first one of the current
busy cycle. In this case the interface immediately starts the transition to
active, so $W^1_1 = T_{\mathrm{W}}$. Therefore
\begin{equation}
  \label{eq:green-frame-1-active}
  W^{\text{green}}_{\text{tandem}}(i) = T_{\mathrm{W}} + s^1_1 - s^1_i \le
  T_{\mathrm{W}} + s^1_i\quad\forall i \ge 0,
\end{equation}
when $b^0$ is active at frame $i$ arrival.

A frame can also reach $b^0$ when all the previous traffic has already
departed. In this case, $b^0$ is either \emph{sleeping} or \emph{transitioning
  to sleep.} In either case, the frame must wait at $b^0$ to end the
transition to sleep and then return to the active state after a time
$T_{\mathrm{W}}$. When the frame finally reaches $b^1$ it must be in the
sleeping state as any pending transition to sleep must have already concluded.
So, the frame just waits at $b^1$ for a time $T_{\mathrm{W}}$ while it is
transitioning to active. So
\begin{equation}
  \label{eq:green-frame-1-allways}
  W^{\text{green}}_{\text{tandem}}(i) = T_{\mathrm{W}}\quad\forall i \ge 0,
\end{equation}
when $b^0$ is sleeping at frame $i$ arrival.

It easily follows that if more energy efficient interfaces are added to the
network in the same fashion, each one will add between $T_{\mathrm{W}}$ and
$T_{\mathrm{W}}+S$ seconds of delay, with $S=\max_i\{s_i\}$. So, if we
consider a series of $n$ energy efficient interfaces, the total delay would be
included in
\begin{equation}
  \label{eq:total-tandem-green}
  \overline W_{\text{first}} + (n-1)T_{\mathrm{W}} \le
  \overline W_{\text{tandem}} \le
  \overline W_{\text{first}} + (n-1)(T_{\mathrm{W}}+S),
\end{equation}
where $\overline W_{\text{first}}$ is the average delay at the first green
interface in the series. Recall that if the first green interface receives
Poisson traffic, $\overline W_{\text{first}} = \overline W_{\text{EEE}}$.
If, however, the first energy efficient interface is in tandem with a previous
non energy efficient interface,
$\overline W_{\text{first}} = \overline W_{\text{tandem}}^{\text{regular}}$.

We can now study the added delay cost due to the energy efficient operation in
a series of tandem switches with identical capacity. We compare the delay of a
series of regular Ethernet interfaces to that of a series of green interfaces.
In the first case, assuming again Poisson arrivals, the total delay is just
the queuing delay at the first interface, as the arrival rate at the rest is
never greater than their output link capacity, plus the one due to the frame
size differences at each successive link, so the total delay is bounded by
$\overline W_{\textrm{M/G/1}} + (n-1)S$. In the energy efficient
tandem, $\overline W_{\text{first}} = \overline W_{\text{EEE}}$, so the
difference is
\begin{equation}
  \label{eq:total-added-delay-tandem}
  \begin{split}
    \Delta \overline W_{\text{tandem}} &=
    \overline W_{\text{EEE}} + (n-1)T_{\mathrm{W}} + (n-1)S \\
    &- (\overline W_{\text{M/G/1}}+ (n-1)S)\\    
    % \frac{T_{\mathrm{W}}}2\left(
    %   1+\frac{T_{\mathrm{W}}}{1+\lambda T_{\mathrm{W}}}
    % \right)
    % +(n-1)T_{\mathrm{W}} \\
    &= \overline W_{\text{tandem}}^{\text{regular}} + (n-1)T_{\mathrm{W}} \\
    &= \frac{T_{\mathrm{W}}}{2}\left(
      2n + \frac{1}{1+\lambda T_{\mathrm{W}}} - 1
    \right)\le nT_{\mathrm{W}}.
  \end{split}  
\end{equation}

As expected, the per interface added delay converges to
\begin{equation}
  \label{eq:per-eee-added-delay-tandem}
  \lim_{n \to \infty} \frac{\Delta \overline W_{\text{tandem}}}{n} = T_{\mathrm{W}}.
\end{equation}

\section{Experimental Validation}
\label{sec:exper-valid}

We have tested our models with the help of the 
ns-2 simulator~\cite{ns-2}. To this end, we have employed an in-house
developed module implementing the EEE frame transmission
policy~\cite{14:_networ_simul-removed}. We chose popular 10GBASE-T interfaces,
so the transition lengths to idle and to active are, respectively,
$2.88\,\mu$s and $4.48\,\mu$s~\cite{802.3az}.
We have performed a series of experiments for each of the developed models
both with synthetic traffic traces and real traffic from the CAIDA
project~\cite{internet15:_caida_ucsd_anony_inter_traces}. Each experiment with
synthetic traffic has been repeated ten times with different random seeds, and
then we have obtained the 95\% confidence interval of every
measure.\footnote{The actual confidence intervals are negligible and not shown
  in the figures to avoid excessive clutter.} The synthetic traffic traces
have been modeled with exponential and Pareto interarrival times, in the
latter case with $\alpha=2.5$.\footnote{Note that Pareto distributions must be
  characterized with a shape parameter $\alpha$ greater than 2 to have finite
  variance. On the other hand, the greater the $\alpha$~parameter, the
  shorter the fluctuations, so a value of $2.5$ is a good trade off to
  have finite variance along with significant fluctuations.} In both cases
frame sizes follow a bimodal distribution to approximate real Internet
traffic~\cite{Murray2012}. We employed a frame size of 100~bytes with a 54\%
probability and a size of 1500~bytes with a 46\% probability.

\subsection{Results for the Traffic Aggregation Scenario}
\label{sec:concentrator-results}

We first proceed with the case where many interfaces send traffic to an
aggregator. The first state of the network consists on a varying number of
interfaces (from 2 to 100), each receiving an independent traffic stream,
although of equal average rate.

\begin{figure}
  \includegraphics[width=\columnwidth]{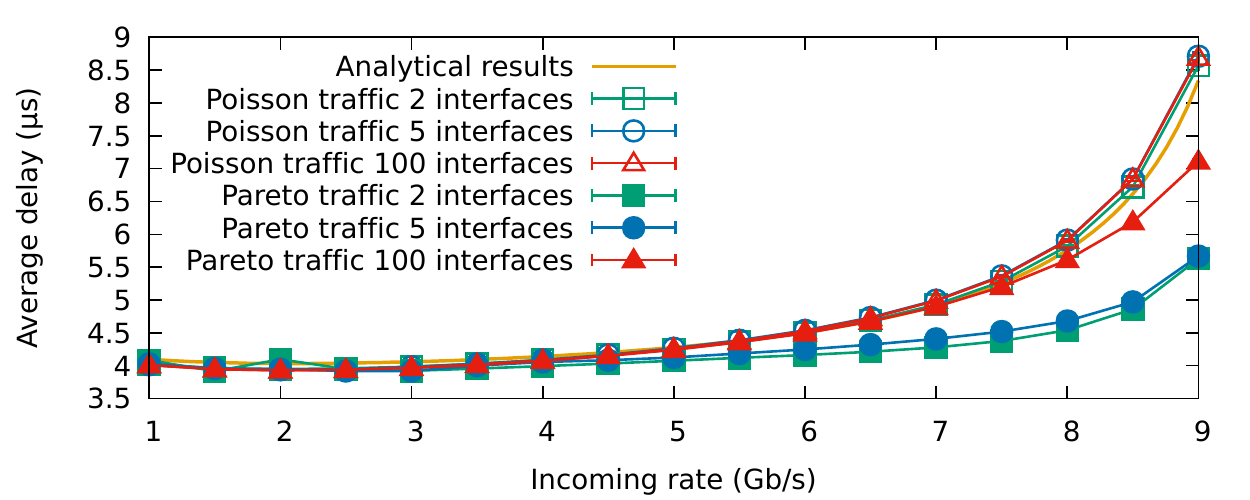}
  \caption{Delay at the second interface when receiving traffic from multiple
    independent non energy efficient interfaces.}
  \label{fig:n-droptail}
\end{figure}  
Figure~\ref{fig:n-droptail} shows the results for the case where only the
second stage interface is energy efficient. As expected, for the scenario with
100 aggregators the results match the model with extraordinary accuracy. Also
note that the results for the scenario with just two interfaces at the first
stage exhibit an unexpected level of accuracy, and only the Pareto traffic
shows little deviations from the theoretical values at the highest loads,
where EEE is less useful, as its energy savings decay with load. Recall that
the model is only valid for large values of contributors and that the output
of each first stage link does not follow a Poisson distribution, since the
frame sizes are not exponentially distributed. As the number of interfaces
increases the results rapidly converge to the model prediction.

\begin{figure}
  \includegraphics[width=\columnwidth]{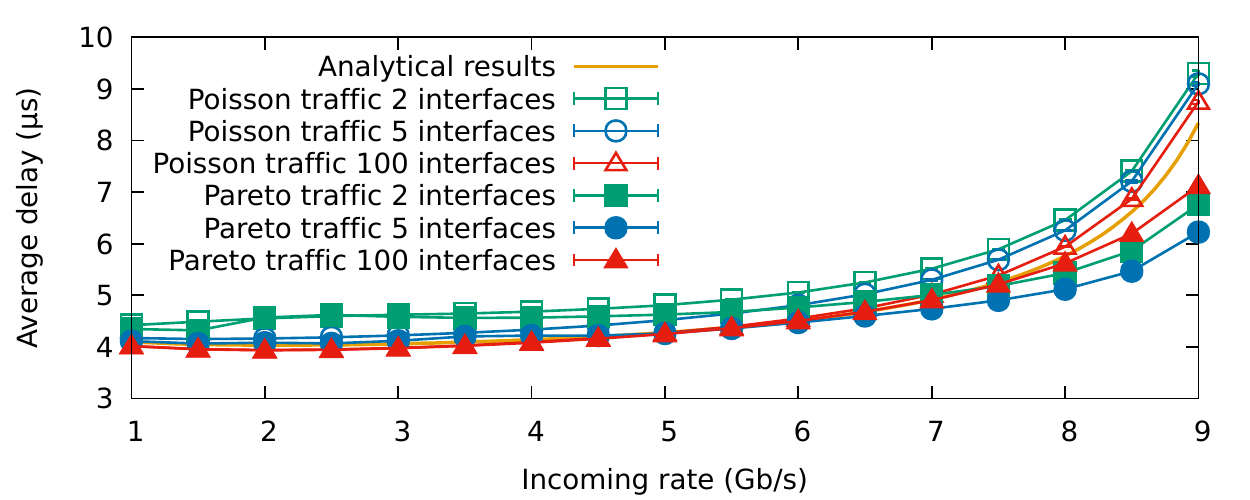}
  \caption{Delay at the second interface when receiving traffic from multiple
    independent energy efficient interfaces.}
    \label{fig:n-greentail}
\end{figure}
The results for the case where all the links have power saving capabilities is
shown in Fig.~\ref{fig:n-greentail}. Again, the results for 100 first-stage interfaces match the model almost perfectly, as expected.
However, in this case the results for the two interfaces case deviate some
more from the model, because the energy saving algorithm adds additional
correlation to the output of the interfaces. As before, results rapidly
converge to the model if the number of interfaces increases, with the results
for just five users showing a very good match. In all cases, the results are
accurate enough to validate the model for any number of aggregating interfaces.

\subsection{Results for the Tandem Network}
\label{sec:tandem-results}

This section shows the results for a network composed of two consecutive
Ethernet interfaces.

\begin{figure}
  \includegraphics[width=\columnwidth]{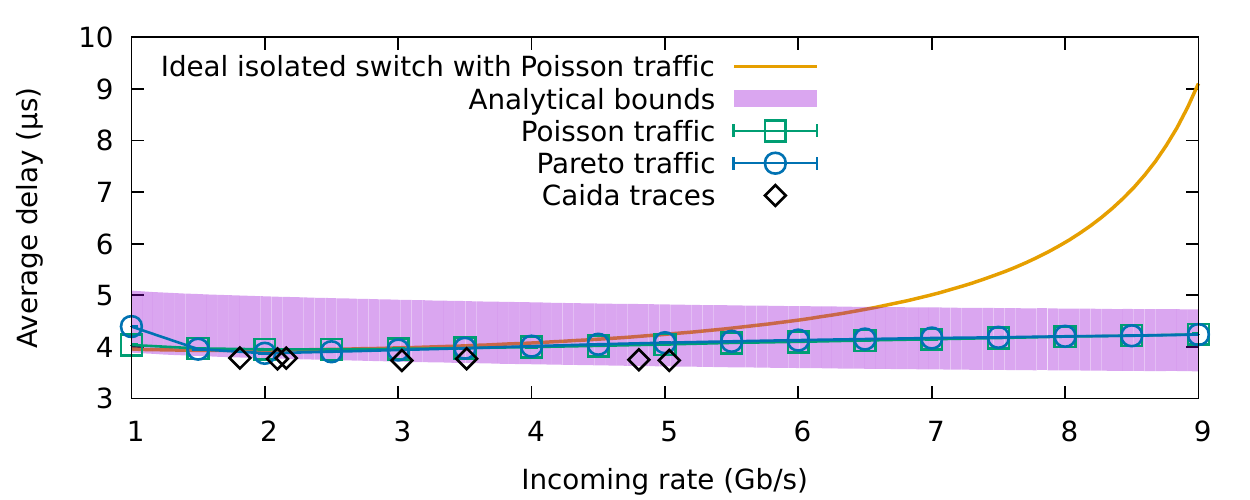}
  \caption{Delay at the second interface when receiving traffic from a single
    regular Ethernet interface.}
  \label{fig:1-droptail}
\end{figure}
The results for the non energy efficient interface followed by an EEE one are
represented in Fig.~\ref{fig:1-droptail}. The delay experienced in the second
interface always falls inside the bounds predicted by the model. Note that,
contrary to what happened in the aggregating scenarios, the delay departs from
that suffered by an isolated link. This is because there is no queuing delay
at this second interface due to traffic load as the incoming traffic rate to
this interface is never greater than its outgoing link capacity. It is also
remarkable that the maximum delay at the second interface monotonically
decreases with the traffic load, with the maximum delay equal to $T_{\mathrm{W}}$
plus the maximum transmission time of a frame, for a very low incoming traffic
rate. The results for the real traffic trace are closer to the minimum value
predicted by the model because its frame length variations are less extreme
than those of our synthetic traces. If all the frame sizes were equal, the
delay would be exactly $T_{\mathrm{W}}$.

\begin{figure}
  \includegraphics[width=\columnwidth]{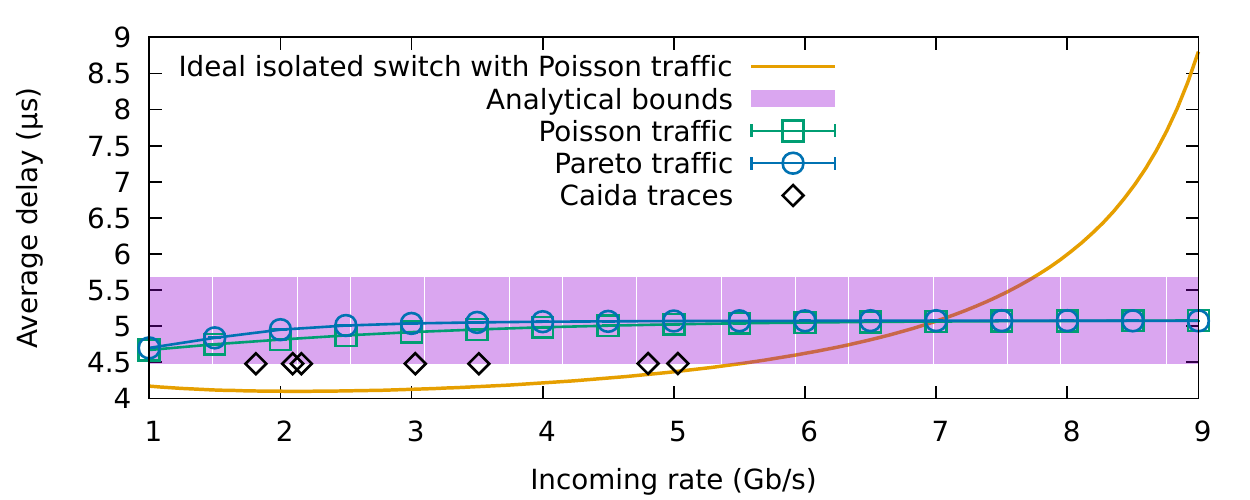}
  \caption{Delay at the second interface when receiving traffic from a single
    power saving interface.}
    \label{fig:1-greentail}
\end{figure}

Figure~\ref{fig:1-greentail} shows the results for the case where both
interfaces are energy efficient. As the model correctly predicts, the delay
bounds at the second (or later) interface are constant, between the setup time
and the setup time plus the transmission of the longest frame. As with the previous case,
there is no queuing delay at the second interface greater than the
transmission time of a single frame.

Finally, we have also tested the performance of tandem networks of variable
lengths. The experimental network consists on a regular interface followed by
up to 100 power efficient interfaces.
\begin{figure}[t]
  \includegraphics[width=\columnwidth]{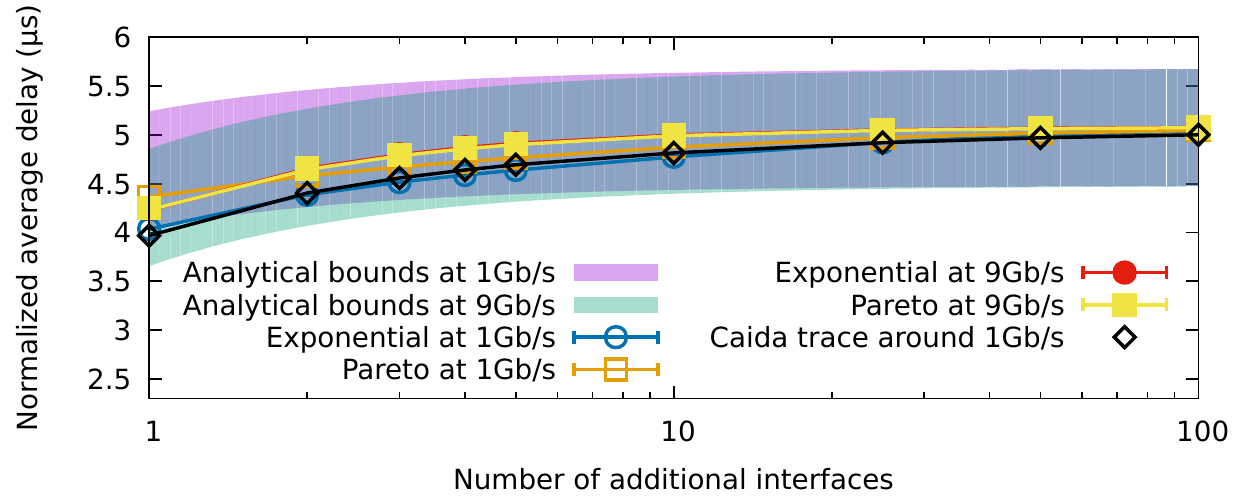}
  \caption{Normalized delay measured on a tandem of $n$ interfaces with
    different kinds of incoming traffic.}
  \label{fig:nsteps-droptail}
\end{figure}
The results in Fig.~\ref{fig:nsteps-droptail} show that the per interface
average delay is between the predicted bounds for different average rates and
all kinds of incoming traffic.

\section{Conclusions}
\label{sec:conclusions}

In this paper we have modeled the delay of traffic traversing a series of EEE
interfaces. Although the delay of a single interface was already well
understood and known to be negligible in most contexts, the study of the
effects of the accumulation had been neglected before.

We have proven and empirically tested that, when the traffic to an energy
efficient link arrives from several previous interfaces, their arrivals can be
modeled with a Poisson process. Therefore, the delay in this scenario can be
modeled like in the isolated case, without regarding whether the traffic first
traversed energy efficient links, regular links or a mix of them. We have also
shown that the delay strictly due to the EEE capabilities of the interface is
never larger than the length of the setup time, usually just a few
microseconds.

Finally, we have also shown how, when the traffic is dominated by a single
contributor, its delay is quite different (and quite smaller) than that of an
isolated interface. The delay in these interfaces never grows larger than the
setup time plus the transmission time of a single frame. This result helps to
calculate budget delays in EEE networks. It is also important
to note that this kind of setup does not add any jitter to the traffic as long
as the frame sizes are constant.

Future work includes extending the current analysis for the case where the
interfaces employ different algorithms to manage the power saving mode, such
as the packet coalescing technique, that waits for several frames to arrive
while in the LPI mode before returning to the active mode.

\bibliographystyle{IEEEtran}
\bibliography{IEEEabrv,library,library-blind}

\end{document}